\documentstyle[preprint,aps,epsf]{revtex}
\tightenlines

\newcommand{\bold}[1]{\mbox{\boldmath $#1$}}	%	bold math character

\begin{document}

%\draft

\title{
Analysis of exchange terms in a projected ERPA Theory
applied to the quasi-elastic 
$(e,e')$ reaction.}

\author{E. Bauer \footnote{Fellow of the Consejo Nacional
de Investigaciones Cient\'{\i}ficas y T\'ecnicas, CONICET.}}
\address{
Departamento de F\'{\i}sica, Facultad de Ciencias Exactas,\\
Universidad Nacional de La Plata, La Plata, 1900, Argentina.}
\author{A. Polls and A. Ramos}
\address{Departament d'Estructura i Constituents de la Mat\`eria,\\
Universitat de Barcelona,
Diagonal 647, E-08028, Spain.}
\date{\today}
\maketitle

\begin{abstract}

A systematic study of the influence of exchange terms in the
longitudinal and transverse nuclear response to quasi-elastic
$(e,e')$ reactions is presented. The study is performed within
the framework of the extended random phase approximation (ERPA),
which in conjuction with a projection method permits a separation of
various contributions tied to different physical processes. The
calculations are performed in nuclear matter up to second order in the
residual interaction for which we take a $(\pi+\rho)$-model with the
addition of the Landau-Migdal $g'$-parameter. Exchange terms are found
to be important only  for the RPA-type
contributions around the quasielastic peak.

\end{abstract}

\section{INTRODUCTION}

The nuclear response to an electromagnetic probe is
a common tool to investigate the behaviour of the atomic nucleus
\cite{kn:book}. At variance with a hadronic probe it allows a
perturbative treatment in the external operator
coupling constant. In this work we will concentrate on
the study of the nuclear response function for
longitudinal and transverse inclusive
quasi-elastic electron scattering reactions. These
responses are experimentally separated
\cite {kn:ba83,kn:me84,kn:me85,kn:bl86,kn:wi95,kn:jo96}, showing that
a simple model like the Fermi gas fails to reproduce the
experimental data. The attempts to go beyond this model can be
classified in two groups. On one side there are the methods that assume
the nucleus as an assembly of non interacting nucleons with
individual           
properties, such as the charge radius, modified with
respect to the vacuum due to the presence of the other
nucleons \cite {kn:no81,kn:do84,kn:ku86,kn:ni92}. Another option
is
to explore the possibilities of a rigorous many-body theory
\cite {kn:ca84,kn:dr87,kn:co88,kn:dr90,kn:fa87,kn:ho80,kn:ch89,kn:pi85,kn:ca91,kn:ca93,kn:bo89,kn:al84,kn:al91,kn:ta88,kn:ta91,kn:ta89,kn:ba91,kn:pace93,kn:ba95,kn:ba96,kn:pol89,kn:chris96,kn:gil97,kn:am92,kn:ji92}
keeping the
nucleons as the essential degrees of freedom with the same properties as
in the vacuum, before resorting to such exotic effects. This work falls
in this second strategy.

Several approaches to the nuclear many body
problem of the nucleus for these processes have been extensively
analyzed in the literature \cite{kn:book}. Microscopic many-body
theories must
deal with short range correlations (SRC) originated from the short range
repulsion of the  nucleon nucleon (NN) interaction. Variational
calculations account for SRC by introducing a Jastrow correlation
factor
explicitly in the wave function. In this way, it is possible to define a
correlated basis function (CBF) and  build a fast
converging perturbation
theory using this basis. There are recent studies of both the
longitudinal and the transversal response for nuclear matter in this
framework \cite{kn:fa87}. Alternatively, the effect of SRC can be incorporated
by introducing a well
behaved effective interaction of G-matrix type or the standard
Landau-Migdal parameterization, with which one can perform perturbation
theory to build other correlations, for instance of RPA
type. As it is not our
aim to describe the full set of approaches, we will comment only on
the ones which lead to our particular theory, which is based on the
perturbative approach.

A simple way to introduce in the response the nucleon-nucleon correlations
originated by
the residual interaction
is by means of the RPA theory, where one particle-one hole
excitations are summed up to infinite order.
Although improving the Fermi gas picture, the RPA approximation is not
able to explain some features of the response as, for instance, the
strength in the ``dip'' region of the transverse response.

An improvement upon the RPA-theory consists of allowing the coupling of
one particle-one hole states to two particle-two
hole ones. This corresponds to what has been called final
state
correlations. Two formalisms study this kind of processes. One is
the second RPA (SRPA) theory (see Ref. \cite{kn:dr90} and
references therein) and the other is the Green
function scheme of Ref. \cite{kn:ho80} (see also
\cite{kn:ch89,kn:pi85,kn:ca91,kn:ca93}). The first one introduces
final
%\cite{kn:ch89}-\cite{kn:ca93}). The first one introduces final
state correlations over the particle-hole bubbles of the
RPA theory. In the second one
the relationship between forward virtual
Compton scattering and inclusive electron scattering
is used to construct a one-body approximation to
quasi-elastic electron scattering. In fact, at large
momentum transfers, where the effect of long range
correlations is negligible, the SRPA and the optical
model Green's function approach should coincide 
\cite{kn:ch89}.

Both the SRPA and the optical-model Green function approaches use
the full residual interaction and allow for many particle-many hole
final states. Still in both approaches the one
body external operator is
limited to create (or destroy) a one particle-one hole pair.
Once the external operator is allowed to scatter a particle
(or hole), then two particle-two hole states stemming from
ground state correlations (GSC) could be activated.
The importance of these GSC
are particularly relevant in the dip region
for the transverse channel \cite{kn:al84}.
A theory to calculate
the response function which takes into account  all
the above requirements is already established
as the Extended RPA (ERPA) theory \cite{kn:ta88,kn:ta89}. 
Still, the application of the
ERPA theory is in general a prohibitively large task.
In Ref. \cite{kn:ba95} we developed a projection method which extracts
the main ingredients of the ERPA theory.
In that work, the scheme was presented and the response was calculated
neglecting the exchange part in the matrix elements of the nuclear
particle-hole interaction. Therefore it seems necessary to complete
the
scheme by investigating and stablishing the influence of the exchange
graphs in the longitudinal and transverse response of
non-relativistic
nuclear matter. Actually, the importance of the
exchange terms in the RPA theory is a well known problem which can not be
satisfactorily solved for finite range interactions, although several
attempts have been made.
In a previous work\cite{kn:ba96}, we developed a simple scheme to evaluate the
full antisymmetric RPA series contributing to the nuclear matter
response, with the result that exchange contributions are important,
specially at low momentum transfer, and can not be accounted for by
simply evaluating the ring (direct) series with an effective $g'$
parameter.

In view of the importance of the RPA exchange terms and the ongoing
efforts to improve the description of the nuclear response it seems
appropriate to explore
whether the exchange terms in the remaining types
of diagrams contained in the ERPA scheme are also important.
In this work, we undertake this task and evaluate for the first time
the contribution of the exchange terms to the nuclear response up to
second order in the effective interaction.

The
paper is organized as follows. In Sect. 2 we present the formalism.
In Sect. 3 the results for the exchange contributions to the
nuclear matter structure function are presented and compared with
the corresponding direct ones.
Finally, the conclusions are summarized in Sect. 4.

\section{FORMALISM}

 The Longitudinal (L) and Transverse (T)
structure functions,
$S_{L,T}({\bf q},\hbar \omega)$, are defined as

\begin{equation}
S_{L,T}(\mbox{\boldmath $q$},\hbar \omega) = -\frac{1}{\pi}\ Im
<0|{{\cal O} _{L,T}}^{\dag} G(\hbar \omega) {\cal O} _{L,T} |0>
\label{eq:uno},
\end{equation}
where $\hbar \omega$ represents the excitation
energy and $q$ the magnitude of the
three momentum transfer. The nuclear ground state
is denoted by $|0>$ while the polarization propagator
$G(\hbar \omega)$ is given by

\begin{equation}
G(\hbar \omega) = \frac{1}{\hbar \omega - H + i \eta}\ -
\frac{1}{\hbar \omega + H - i \eta}\
\label{eq:green},
\end{equation}
where $H$ is the nuclear Hamiltonian. Explicit forms
for the external excitation operators
${\cal O} _{L,T}$ are given by

\begin{equation}
{\cal O}_L = \sum_{j=1}^{A} \frac{1 + \tau_3 (j)}{2}
e^{i  \mbox{\boldmath $q$}\cdot \mbox{\boldmath $x$}_j}
\label{eq:opl},
\end{equation}

\begin{equation}
{\cal O}_T = \frac{1}{2 m q}
\sum_{j=1}^{A} \{ \frac{1 + \tau_3 (j)}{2}
[ \mbox{\boldmath $q$} \times \{ \mbox{\boldmath $p$}_j ,
e^{i \mbox{\boldmath $q$} \cdot \mbox{\boldmath $x$}_j } \}] +
i \frac{\mu_s + \mu_v \tau_3 (j)}{2}
[\mbox{\boldmath $q$}\times (\mbox{\boldmath $\sigma$} (j) \times 
\mbox{\boldmath $q$})]
e^{i \mbox{\boldmath $q$} \cdot \mbox{\boldmath $x$}_j} \}
\label{eq:opt},
\end{equation}
where $m$ is the nucleonic mass, {\boldmath $x_j$}
and {\boldmath $p_j$} denote the 
position and momentum operators for
individual nucleons, and
 $\mu_s = 0.88$, $\mu_v = 4.70$ are related to the proton and
neutron magnetic moments. In fact, ${\cal O} _L$ is the charge
density operator while
${\cal O}_T$ is
related to the convection and magnetization
current density. The structure functions are related
 to the response functions,
$R_{L,T}(\mbox{\boldmath $q$},\hbar \omega)$,
through the usual dipole electromagnetic
form factor $G_E(q,\hbar \omega)$:
\begin{equation}
G_E (q,\hbar \omega) = [ 1 +
\frac{ (\hbar c q)^2 - (\hbar \omega)^2}{(839 {\rm MeV})^2}]^{-2}
\label{eq:elcf}.
\end{equation}

We introduce now the usual projection operator $P$,
which projects into $npnh$ configurations
with $n$ = 0, 1, defined respect to the HF vacuum, which corresponds
to the case $n=0$ and is denoted by $|~>$.
In addition, two projection operators $Q$ and $R$ are introduced. The
action of $Q$ ($R$) is to project onto
the $npnh$ space with $n$ being an even integer
greater or equal to two ($n$ odd greater
or equal to three). Explicit expressions are given
by

\begin{equation}
P = \sum_{ \begin{array}{c}
	     n = 0,1
	   \end{array} }
	    |n><n|
\label{eq:prop},
\end{equation}

\begin{equation}
Q = \sum_{ \begin{array}{c}
	     \makebox{$n$ even} \\ n \geq 2
	   \end{array} }
	    |n><n|
\label{eq:proq},
\end{equation}

\begin{equation}
R = \sum_{ \begin{array}{c}
	     \makebox{$n$ odd} \\ n \geq 3
	   \end{array} }
	    |n><n|    \ ,
\label{eq:pror}
\end{equation}
where $|n>$ indicates a $np nh$ configuration.

In the literature
only one projection operator,
which is the sum of $Q$ and $R$, is usually
used. The present separation is done for convenience
as it helps to clarify the role of $3p3h$ configurations
(see Ref. \cite{kn:ba95}).
It is easy to
verify that $P + Q + R = 1$, $P^2 = P$,
$Q^2 = Q$, $R^2 = R$ and $PQ = QP =
PR = RP = QR = RQ = 0$.

Inserting the identity in Eq. (\ref{eq:uno})
one obtains

\begin{equation}
S = S_{PP} + S_{PQ} + S_{QP} + S_{QQ} + S_{PR} +
S_{RP} + S_{RR} + S_{QR} + S_{RQ}
\label{eq:dos} \ ,
\end{equation}
where, for simplicity, we have omitted the
subscripts $L$, $T$. The expression for
$S_{PP}$ is given by

\begin{equation}
S_{PP}(\mbox{\boldmath $q$},\hbar \omega) = -\frac{Im}{\pi}\
<0|{\cal O}^{\dag} P G(\hbar \omega) P {\cal O} |0>
\label{eq:sppt};
\end{equation}
and similar expressions can be written for
$S_{PQ}$, etc. To evaluate the propagators
$PGP$, $PGQ$, etc., one has to solve
 the following equation:

\begin{equation}
{\cal G} \cdot {\cal G}^{-1} = I
\label{eq:ide},
\end{equation}
where

\[ {\cal G} = \left( \begin{array}{ccc}
	 PGP & PGQ & PGR \\
	 QGP & QGQ & QGR \\
	 RGP & RGQ & RGR
	       \end{array} \right) \makebox[.6in][l]{,}
    I= \left( \begin{array}{ccc}
	 P & 0 & 0 \\
	 0 & Q & 0 \\
	 0 & 0 & R
	       \end{array} \right). \]

This is an easy task once the properties of the
projection operators are employed. Keeping terms up to
second order in the nuclear interaction, all $3p3h$ final
state contributions cancel each other\cite{kn:ba95}.
Thus, from all terms of Eq.
(\ref{eq:dos}), only
$S_{PP}$, $S_{QP}$ ($S_{PQ}$) and $S_{QQ}$ survive:

\begin{eqnarray}
S_{PP} & = & -\frac{Im}{\pi}\
<0|{\cal O}^{\dag} P
\frac{1}{\hbar \omega - H - \Sigma^{PQP} - Re \Sigma^{PRP}
+ i \eta}\
P {\cal O} |0>, \label{eq:spp2} \\ [7.mm]
S_{QP} & = & -\frac{Im}{\pi}\
<0|{\cal O}^{\dag}
P \frac{1}{\hbar \omega - H_0 + i \eta}\ P
H_{res} Q
\frac{1}{\hbar \omega - H_0 + i \eta}\
Q {\cal O} |0>, \label{eq:sqp2} 
\end{eqnarray}
and
\begin{equation}
S_{QQ}  =  -\frac{Im}{\pi}\
<0|{\cal O}^{\dag} Q
\frac{1}{\hbar \omega - H_0 + i \eta}\
Q {\cal O} |0>, \label{eq:sqq2}
\end{equation}
where $S_{QP}$ equals $S_{PQ}$. The self energy
operators introduced in Eq. (\ref{eq:spp2}) are given by

\begin{eqnarray}
\Sigma^{PQP} & = & P H_{res} Q \frac{1}{\hbar \omega -H_0 + i \eta}\
Q H_{res} P
\label{eq:selq}
\end{eqnarray}
and

\begin{eqnarray}
Re \Sigma^{PRP} & =  & - P H_{res} R \frac{{\cal P}}{\hbar \omega -H_0
}\ R H_{res} P
\label{eq:selr},
\end{eqnarray}
where ${\cal P}$ denotes the principal value.
We have separated the total Hamiltonian $H$ into a one-body
part, $H_0$, and a residual interaction, $H_{res}$.

As pointed out in Ref. \cite{kn:ba95}, there
is still a contribution stemming from a $3p3h$ configuration given by
the real part of $3p3h$-self-energy insertion
($Re \Sigma^{PRP}$) in Eq. (\ref{eq:spp2}).
That is, up to second order, no
$3p3h$ physical state is possible, but virtual intermediate $3p3h$
configurations produce a shift in the ground state energy
(see Ref. \cite{kn:ba95} and Ref. \cite{kn:ta91} for
more details).
The next step is to establish the structure
of the ground state. Including the
ground state correlations perturbatively
one gets, up to first order in
the residual interaction,
\begin{equation}
|0> = |> - {H_0}^{-1} Q H_{res} P |>
\label{eq:cgs},
\end{equation}
with $|>$ being the Hartree Fock ground state.
More explicitly, one can write
\begin{equation}
|0> = |> - \sum_2 \frac{<2|H_{res}|>}{\varepsilon_{gsc}}\ |2>
\label{eq:cgse},
\end{equation}
where the quantity $\varepsilon_{gsc}$ refers to the energy
of the first order correction to the ground state energy.

The aim of this
section is to present the formalism showing explicitly antisymmetric
matrix elements. The guidelines to obtain analytical expressions
are given in Ref. \cite{kn:ba95},
where direct contributions were studied. To complete
the scheme, expressions for the exchange
self-energy insertions
are given in Appendix A and exchange
terms to the structure function are presented in Appendix B.
Also, and in order to simplify the calculation, we will limit
ourselves to the case in which the external operator is attached
to the same bubble. We will study some exceptions to this
as a consequence of antisymmetrization.
The three non vanishing contributions, $S_{PP}$, $S_{QP}$
and $S_{QQ}$, will be analyzed separately below. Special
attention will be paid to $S_{PP}$ as its structure
is very rich and represents the main contribution to the
response function.

\vspace{3 mm}

\subsection{$S_{PP}$ contribution:}

Let us carefully analyze all graphs stemming from $S_{PP}$.
To do this,
we first insert the definition of $P$ given
by Eq. (\ref{eq:prop}) into
$S_{PP}$ [Eq. (\ref{eq:spp2})]:

\begin{eqnarray}
S_{PP} = -\frac{1}{\pi}\ Im
\sum_{n, n' = 1.}
<0|{\cal O}^{\dag} |n>
<n|
\frac{1}{\hbar \omega - H_0 - H_{res} -
\Sigma^{PQP} - Re \Sigma^{PRP} + i \eta}\
\nonumber \\ [5.mm]
|n'><n'| {\cal O} |0>
\label{eq:spps}.
\end{eqnarray}

Using the ground state given by Eq. (\ref{eq:cgse})
in the expression and neglecting all third and higher orders terms
except the ones with self
energy insertions,  one can write
\begin{equation}
S_{PP}={S_{PP}}^{Lindhard+self-energy} +
       {S_{PP}}^{first \, order \, RPA} +
       {S_{PP}}^{second \, order \, RPA}
\label{eq:sppsep},
\end{equation}
where
\begin{eqnarray}
{S_{PP}}^{Lindhard+self-energy} = -\frac{1}{\pi}\ Im
\sum_{1, 1'}
<|{\cal O}^{\dag} |1>
<1|
\frac{1}{\hbar \omega - H_0 -
\Sigma^{PQP} - Re \Sigma^{PRP} + i \eta}\
\nonumber \\ [5.mm]
|1'>_{\it a} <1'| {\cal O} |> \ ,
\label{eq:spplise}
\end{eqnarray}

\begin{eqnarray}
{S_{PP}}^{first \, order \, RPA} & = & -\frac{1}{\pi}\ Im \{
\sum_{1, 1'}
<|{\cal O}^{\dag} |1>
\frac{1}{\hbar \omega - \varepsilon_{1} + i \eta}\
<1| H_{res} |1'>_{\it a}
\nonumber \\ [5.mm]
& & \frac{1}{\hbar \omega - \varepsilon_{1'} + i \eta}\
<1'| {\cal O} |>
\nonumber \\ [5.mm]
& & - 2 \sum_{1} \sum_{2}
\frac{<| H_{res} |2>_{\it a}}{\varepsilon_{gsc}}\
<2|{\cal O}^{\dag} |1>
\frac{1}{\hbar \omega - \varepsilon_{1} + i \eta}\
<1| {\cal O} |> \} \ ,
\label{eq:sppfrpa}
\end{eqnarray}

\begin{eqnarray}
{S_{PP}}^{second \, order \, RPA} & = & -\frac{1}{\pi}\ Im \{
\sum_{1, 1', 1''}
<|{\cal O}^{\dag} |1>
\frac{1}{\hbar \omega - \varepsilon_{1} + i \eta}\
<1| H_{res} |1'>_{\it a}
\nonumber \\ [5.mm]
& & \frac{1}{\hbar \omega - \varepsilon_{1'} + i \eta}\
<1'| H_{res} |1''>_{\it a}
\nonumber \\ [5.mm]
& & \frac{1}{\hbar \omega - \varepsilon_{1''} + i \eta}\
<1''| {\cal O} |>
\nonumber \\ [5.mm]
& & - 2 \sum_{1, 1''} \sum_{2}
\frac{<| H_{res} |2>_{\it a}}{\varepsilon_{gsc}}\
<2|{\cal O}^{\dag} |1>
\nonumber \\ [5.mm]
& & \frac{1}{\hbar \omega - \varepsilon_{1} + i \eta}\
<1| H_{res} |1'>_{\it a}
\frac{1}{\hbar \omega - \varepsilon_{1'} + i \eta}\
<1'| {\cal O} |>
\nonumber \\ [5.mm]
& & + \sum_{1} \sum_{2,2'}
\frac{<| H_{res} |2>_{\it a}}{\varepsilon_{gsc}}\
<2|{\cal O}^{\dag} |1>
\nonumber \\ [5.mm]
& & \frac{1}{\hbar \omega - \varepsilon_{1} + i \eta}\
<1|{\cal O}|2'>
\frac{<2| H_{res} |>_{\it a}}{\varepsilon_{gsc}}\ \} \ ,
\label{eq:sppsrpa}
\end{eqnarray}
and $H_0|n>={\varepsilon_{n}}|n>$.

{}From Eqs. (\ref{eq:selq}) and (\ref{eq:selr})
the self-energy insertions read now
\begin{equation}
<1|\Sigma^{PQP}|1'> =
\sum_{2}
<1| H_{res} |2>_{\it a}
\frac{1}{\hbar \omega - {\varepsilon_{2}} + i \eta}\
<2| H_{res} |1'>_{\it a} \ ,
\label{eq:selfpqp}
\end{equation}
and
\begin{equation}
<1|Re \Sigma^{PRP}|1'> =
- \sum_{3}
<1| H_{res} |3>_{\it a}
\frac{{\cal P}}{\hbar \omega - {\varepsilon_{3}} }\
<3| H_{res} |1'>_{\it a} \ .
\label{eq:selfprp}
\end{equation}
In all expressions
we have made explicit indication of antisymmetric matrix
elements.

In Fig. 1 some graphs contributing to
$S_{PP}$ are shown. Within brackets we have collected direct
plus exchange contributions. Let us start by analyzing
the contribution to ${S_{PP}}^{Lindhard+self-energy}$.
The presence of self energy operators makes the energy
denominators in the r.h.s. of Eq. (\ref{eq:spplise}) to be
non diagonal in our particle-hole basis. Non diagonal terms,
shown by graphs $SE2D$, $SE2E$ and $SE2E'$
in Fig. 1, are evaluated at second
order. For diagonal ones, shown by graphs
$SE1D$, $SE1E$, $SE3D$ and $SE3E$ in Fig. 1,
we first build up an antisymmetric self-energy insertion
and then sum it up to infinite order.

The first two orders leading to the RPA response
(eqs. (\ref{eq:sppfrpa}) and (\ref{eq:sppsrpa})), are
shown by graphs $RPA1D$ to $RPA2E'$ in Fig. 1, where
only the forward going contributions are explicitly shown, that is,
the ones stemming from the first terms on the r.h.s.
of the above mentioned equations.
If exchange
terms were neglected one would be able to sum terms up to infinite
order leading to the usual ring series.

As mentioned, in this
paper we keep terms up to second order in the
evaluation of exchange contributions. However, for the RPA type
diagrams we use the method
described in Ref. \cite{kn:ba96}, which allows to effectively sum up the full
antisymmetric RPA series. That method is based on
splitting the interaction into a pure contact
part and a remaining part chosen such that the second-order ring coincides
with the full ring series. The pure contact term allows a straightforward
evaluation of the antisymmetric RPA series up to infinite order, while
only terms up to second
order are retained for the remaining part of the interaction.

\vspace{3 mm}

\subsection{$S_{QP}$ contribution:}

Using the definition of $P$ and $Q$ into $S_{QP}$
[Eq. (\ref{eq:sqp2})] we have

\begin{eqnarray}
S_{QP} & = & - 2 \frac{1}{\pi}\ Im \{
\sum_{1, 2, 2'}
<|{\cal O}^{\dag} |1>
\frac{1}{\hbar \omega - \varepsilon_{1} + i \eta}\
<1| H_{res} |2>_{\it a}
\nonumber \\ [5.mm]
& & \frac{1}{\hbar \omega - \varepsilon_{2} + i \eta}\
<2| {\cal O} |2'>
\frac{<2'| H_{res} |>_{\it a}}{\varepsilon_{gsc}} \} \ .
\label{eq:sqp3}
\end{eqnarray}
Note that as ${\cal O}$ is a one body operator, it
can scatter a particle (or hole) or create (or
destroy) a particle-hole pair. That is, the Hartree-Fock
ground state is not connected to a $2p2h$ configuration
through ${\cal O}$.

In Fig. 2 we present the second order contributions to
$S_{QP}$. As a consequence of antisymmetrization, a direct
term where the external operator is attached to a different
bubble has come into play (given by graph $SQPD'$
of this figure). 
Naturally, when we act with the antisymmetrization
operator over this term, the same set of graphs
appears. We account for this through a factor two.

\vspace{3 mm}

\subsection{$S_{QQ}$ contribution:}

Finally, the expression for $S_{QQ}$
[Eq. (\ref{eq:sqq2})] is simply

\begin{eqnarray}
S_{QQ} & = & - \frac{1}{\pi}\ Im \{
\sum_{2, 2', 2''}
\frac{<| H_{res} |2>_{\it a}}{\varepsilon_{gsc}}\
<2|{\cal O}^{\dag} |2'>
\frac{1}{\hbar \omega - \varepsilon_{2'} + i \eta}\
<2'|{\cal O}|2''>
\nonumber \\ [5.mm]
& & \frac{<2''| H_{res} |>_{\it a}}{\varepsilon_{gsc}}\} \ .
\label{eq:sqq3}
\end{eqnarray}
Here, the only possible action for the
external operator is to scatter a particle
or a hole.

In Fig. 3, we present the main contributions to $S_{QQ}$.
Graphs $SQQ3D'$ presents a direct contribution with the external
operator attached to a different bubble, in complete analogy
with $S_{QP}$.

\section{RESULTS FOR NUCLEAR MATTER}

In order
to benefit from the advantage of translational
invariance, results have been obtained for infinite
nuclear matter at normal saturation density corresponding to a Fermi
momentum $k_F=1.36$ {}fm$^{-1}$.

For the residual interaction $H_{res}$ we
assume the ( $\pi$ + $\rho$ )-exchange model
at the static limit with the addition of the Landau
Migdal $g'$-parameter. In pionic units it reads

\begin{equation}
H_{res}(l) = \frac{f_{\pi}^2}
{\mu_{\pi}^2}
\Gamma_{\pi}^2 (l)
(\tilde{g}' (l) \mbox{\boldmath $\tau \cdot \tau ' $} \,
\mbox{\boldmath $\sigma \cdot \sigma ' $} +
\tilde{h}' (l) \mbox{\boldmath $\tau \cdot \tau ' $} \,
\mbox{\boldmath $\sigma \cdot \widehat{l} $} \,
\mbox{\boldmath $\sigma ' \cdot \widehat{l} $} ) \ ,
\label{eq:inter}
\end{equation}
with

\begin{equation}
\tilde{g}' (l) = g' -
\frac{\Gamma_{\rho}^2 (l)}
{\Gamma_{\pi}^2 (l)}\ C_{\rho}
\frac{l^2}{l^2 + \mu_{\rho}^2} \
\label{eq:gpt},
\end{equation}

\begin{equation}
\tilde{h}' (l) =
- \frac{l^2}{l^2 + \mu_{\pi}^2} \ +
\frac{\Gamma_{\rho}^2 (l)}
{\Gamma_{\pi}^2 (l)}\ C_{\rho}
\frac{l^2}{l^2 + \mu_{\rho}^2} \ ,
\label{eq:hpt}
\end{equation}
where $\mu_{\pi} \hbar c$ ($\mu_{\rho} \hbar c$ )
is the pion (rho) rest mass and $C_{\rho} = 2.18$.
For the form factor of the
$\pi NN$ ($\rho NN$) vertex
we have taken

\begin{equation}
\Gamma_{\pi, \rho} (l) =
\frac{\Lambda_{\pi, \rho}^2 - (\mu_{\pi, \rho} \hbar c)^2 }
{\Lambda_{\pi, \rho}^2 + (\hbar c l)^2} \ ,
\label{eq:ver}
\end{equation}
with $\Lambda_{\pi} = 1.3$ GeV and $\Lambda_{\rho} = 2.$
GeV.
The role of the $g'$-parameter is to account for short
range correlations.
Note that for a pure contact interaction
exchange contributions have been traditionally included in the RPA series
by a 
redefinition of the Landau-Migdal parameters.
In particular, standard $g'$-values range from
0.7 to 0.95 but, when redefined to account for antisymmetric
terms, the values are lowered and range
from 0.5 to 0.7 \cite{kn:sh89}.
As we evaluate explicitly exchange graphs a standard
$g'$-value from 0.7 to 0.95 should be used.
We have employed
the value $g'=0.7$. In addition, in all the diagrams considered
in our calculations, the nucleon lines have been dressed in 
an average way by taking a  momentum independent effective mass
value of $m^*/m=0.85$.

{}From Eqs. (\ref{eq:spplise})-(\ref{eq:sqq3}), explicit
expressions for the structure functions in nuclear
matter can be  obtained, where the sums over the different
configurations are replaced by multidimensional integrals.
The expressions of the direct terms were reported in Ref.
\cite{kn:ba95} and will not be repeated here.
In Appendix A we give the
exchange self-energy insertions appearing in Eq. (\ref{eq:spplise}) (see
Eqs. (\ref{eq:selfpqp}) and (\ref{eq:selfprp})), while
in Appendix B the exchange contributions to Eqs.
(\ref{eq:spplise})-(\ref{eq:sppsrpa}) and
(\ref{eq:sqp3})-(\ref{eq:sqq3}) are shown (see also Figs. 1-3).
The multiple integrations have been performed using a Monte Carlo
technique.

Let us analyse the three non-vanishing contributions,
$S_{PP}$, $S_{QP}$ and $S_{QQ}$, to the response.
We follow the notation already shown
in Figs. 1-3.

Tables I and II give the results for the $S_{PP}$ channel.
In Table I we compare all direct
and exchange contributions from self-energy insertions.
To avoid divergencies, diagonal self-energy insertions are evaluated up to
infinite order \cite{kn:al91}. To do this, an average over the hole
momentum of the bubble where the self-energy is
attached should be done. This procedure is outlined in
Appendix A.
{}From the table it is clear that, although small in general, the exchange
diagrams can amount to a non-negligible fraction of the direct ones,
especially at energies around and below the quasiparticle peak.
This is also visualized in Fig. 4, where the structure function
including only the direct self-energy diagrams (long-dashed line) is compared
to that containing, in addition, the exchange ones (full
line) for both the longitudinal (upper part) and the transverse (lower
part) channels. The short-dashed  line is the
free structure function calculated with an effective mass of value
of $m^*/m=0.85$.
As has been observed before \cite{kn:fa87,kn:pol89,kn:chris96,kn:gil97}, the
dressing of the nucleon lines by self-energy insertions smears out the structure function, moving
strength out from the quasiparticle peak to the high missing energy region.

Notice that in the diagrams shown in Figs. 1-3 we have not explicitly included
the first order self-energy insertion on the fermion propagators. Instead of
this, we have preferred to use a Lindhard function calculated with an
effective mass $m^*$ in the nucleon propagators and, therefore, it already contains in an average way the effects of those self-energy insertions propagated
to
all orders. Actually, the use of $m^*$ is equivalent to having a real
 and energy independent self-energy parameterized by a function
quadratic in the momentum  of
the nucleon. These mean field single-particle states define the HF ground state
and the basis in which the perturbation theory has been
constructed.

 The second order self-energy diagrams SE1D,SE3D,SE1E,SE3E are responsible for
the appearance of an imaginary part which yields a width to the nucleon lines,
which in turn is the responsible for the observed spreading in the nuclear 
response. 

In Table II we analyse the RPA-type correlations showing
explicit results for the first and second order contributions.
This is done for the transverse channel as RPA-type
correlations are zero for the longitudinal one
due to the election of our interaction with no $f$ or $f'$ Landau-Migdal
terms. In the last column we also
present the results for RPA correlations when exchange
contributions are included up to infinite order, following the scheme of
Ref. \cite{kn:ba96}.
{}From Table II is clear that exchange terms of RPA type are very important.
Their size is comparable (even bigger) to other direct
diagrammatic contributions
to the structure function and, therefore, they should not be neglected.
Given the magnitude of these RPA exchange terms, the differences
between the next-to-last and last columns of Table II also suggest that it is
important to
sum them up to infinite order as was done in Ref. \cite{kn:ba96}.
The effect of the RPA diagrams in the transverse structure function
is also displayed in Fig. 5, where the full antisymmetric
RPA series (full line)
is compared to the direct ring series (long-dashed line). We observe
that the transfer
of strength from the low to the high energy region typical of the
polarization (ring) diagrams is partly restored by the incorporation of
the exchange diagrams.

In Table III, we study the $S_{QP}$ channel. In this kind of graphs
the external operator creates (or destroys) a particle-hole
pair and scatters a particle (or a hole). We have evaluated
the case where the external operator scatters a particle. The
case where it scatters a hole is negligible as can be
found in Ref. \cite{kn:ba95}. As mentioned above, due to the
action of the antisymmetrization operator a graph where the
external operator is attached to different bubbles has to be
considered. This graph is $SQPD'$ in Fig. 2 and has some
influence as can be seen from Table III. The other exchange graphs
are negligible.

In Table IV the results for the $S_{QQ}$ channel are shown. As for the
$S_{QP}$ channel, exchange graphs are very small.
The importance of the ground state correlation diagrams is clearly
seen in Fig. 6, where the full lines represent the addition of
$S_{QP} + S_{QQ}$
to the long-dashed lines, which contain the
$S_{PP}$ contributions. In the longitudinal channel (upper part), only
the self-energy terms contribute to $S_{PP}$, while the transverse
channel (lower part) contains, in addition, the RPA-type correlations.
In view of these results it is clear that incorporating the $S_{QP}$ and
$S_{QQ}$ channels is necessary in any perturbative calculation of the
nuclear matter response as observed in Ref.\cite{kn:ba95}.
 Having established in this work the smallness of
the corresponding exchange terms is particulary interesting, since
the calculation of these channels can be restricted to the direct
graphs thus avoiding a great deal of numerical computations.

In addition our findings also support the idea that the use of an 
effective $g'$ to account for exchange terms in the nuclear response is 
not appropriate. This point was already raised in Ref. \cite{kn:ba96},
where we built a prescription to calculate the full antisymmetric RPA series
of the nuclear response. In that work, we showed that the use of a 
standard average prescription for $g'$   was not able to reproduce the RPA
antisymmetric
response, especially for intermediate values of the momentum transfer.
Using the average $g'$ for calculating the other types of correlations would
not be appropriate either because we have shown that they are basically 
dominated by the direct contributions. This is visualized in Table V where, 
changing
the $g'$  parameter to an effective value of $g'=0.5$ does not, in the first
place,
reproduce the antisymmetric RPA response contained implicitly in the first column 
of the transverse part (a point already raised in Ref.
\cite{kn:ba96})
 but 
also induces non negligible modifications in the other contributions, especially those related to ground state correlations (compare the next-to-last and last
columns).

{}From our study we conclude that the nuclear response is basically
dominated by
the direct diagrams and the exchange ones only need to be considered
 (and to all orders) for the RPA type correlations.

\section{CONCLUSIONS}

A projection method to extract the main contributions 
of the ERPA theory with the explicit inclusion of
exchange terms has been  developed. This work is a continuation of
a previous one \cite{kn:ba95} in which only direct terms were studied.
Here we have tried to clarify the
importance of exchange terms of the particle-hole interaction, by performing
a quantitative analysis of their influence in the nuclear response.

The projection method classifies the contributions to the nuclear
structure function into three channels, called $S_{PP}$,
$S_{QP}$ and $S_{QQ}$ with $P$ and $Q$ being projections
operators defined in Sect. 1. This separation permits to study the effects
of the different types of correlations. 
In this sense, $S_{PP}$ represents
final state correlations, $S_{QQ}$ ground state correlations
and $S_{QP}$ the interference between them. Through the
analysis of our results we can conclude that all types of correlations
are important and should be considered when one
studies the nuclear response.

After this statement, the 
the problem is the big number of graphs
which should be evaluated when the exchange part of the
nucleon-nucleon interaction is retained. Before the numerical calculation
we can see no reason to neglect any contribution.
Our calculations show that for final state correlations,
i.e. self-energy insertions and mainly RPA-type correlations, 
the exchange graphs are relevant 
in agreement with Refs.
\cite{kn:sh89,kn:bu90}.
On the other hand, they can be neglected
for  ground state correlations.
Also, within the energy-momentum region under
consideration, exchange terms can not be parameterized
by a redefinition of the $g'$ parameter.

It is also important to stress that the interaction employed
and in particular the value for $g'$, comes from parameterizations
of processes
at a lower energy-momentum region than the ones considered here.
Those values do not
necessarily hold for us. Also from Table V, we see that a small
change in one parameter can produce a noticeable change in
the structure function. In any case, our scope was not a search of
the optimal paremeters that produce a good agreement with the experimental
data but a careful analysis of the exchange diagrams and
our conclusions should
remain valid in a wide variety of situations. 

In summary,
our study shows
that the nuclear response is basically dominated by the direct diagrams,
the most relevant being those of Figs. 1-3, and the exchange
contributions
 only need to be considered (and to all orders) for the RPA type
 correlations, which can be evaluated using the prescription of Ref.
\cite{kn:ba96}.

\newpage

\section*{APPENDIX A}

In this appendix we show explicit expressions for
exchange contributions to the self-energy insertions
Direct contributions
can be found in Ref. \cite{kn:ba95}.
Exchange self-energy insertions from
eqs. (\ref{eq:selfpqp}) and (\ref{eq:selfprp}) (which
contribute to graphs $SE1E$
and $SE3E$ of Fig. 1, respectively), are given by

\begin{eqnarray}
(\Sigma^{PQP} (\bold{Q}, \nu, \bold{h}))^{part. \, exch.} & = &
- \frac{1}{(2 \pi)^4}\
(\frac{f_{\pi}^2}{4 \pi \hbar c} \ )^2
\frac{m c^2 \, {k_F}^4}{\mu_{\pi}^4} \ \int d^3 k \ \int d^3 k' \
\theta (| \bold{h} + \bold{Q} - \bold{k} | - 1)
\nonumber \\ [5.mm]
& &
\theta (1 - | \bold{h} + \bold{Q} - \bold{k} - \bold{k'} | )
\theta (| \bold{h} + \bold{Q} - \bold{k'} | - 1)
\Gamma_{\pi}^2 (k) \Gamma_{\pi}^2 (k')
\nonumber \\ [5.mm]
& &
(3 \tilde{g}'^2 - ( 2 (\bold{\widehat{k}.\widehat{k'}} )^2 - 1)
 \tilde{h}'^2  + 2 \tilde{g}' \tilde{h}' )
\nonumber \\ [5.mm]
& &
\frac{1}
{\nu - ( Q^2/2 + \bold{h.Q} - \bold{k.k'} ) + i \eta }  \
\label{eq:sigq}
\end{eqnarray}
and

\begin{eqnarray}
(Re \, \Sigma^{PRP} (\bold{Q}, \nu, \bold{h}))^{part. \, exch.} & = &
\frac{1}{(2 \pi)^4}\
(\frac{f_{\pi}^2}{4 \pi \hbar c} \ )^2
\frac{m c^2 \, {k_F}^4}{\mu_{\pi}^4} \ \int d^3 k \ \int d^3 k' \
\theta (1 - | \bold{h} + \bold{Q} - \bold{k} | )
\nonumber \\ [5.mm]
& &
\theta ( | \bold{h} + \bold{Q} - \bold{k} - \bold{k'} | - 1 )
\theta (1 - | \bold{h} + \bold{Q} - \bold{k'} | )
\Gamma_{\pi}^2 (k) \Gamma_{\pi}^2 (k')
\nonumber \\ [5.mm]
& &
(3 \tilde{g}'^2 - ( 2 (\bold{\widehat{k}.\widehat{k'}} )^2 - 1)
 \tilde{h}'^2  + 2 \tilde{g}' \tilde{h}' )
\nonumber \\ [5.mm]
& &
\frac{1}
{\nu - ( Q^2/2 + \bold{h.Q} + \bold{k.k'} ) }  \ .
\label{eq:sigr}
\end{eqnarray}
We have used dimensionless quantities $\bold{Q=q}/k_F$
and $\nu=\hbar \omega/ 2 \varepsilon_F$;
$k_F$ and $\varepsilon_F$ being the Fermi
momentum and energy, respectively.

In order to simplify the calculation, it is a good approximation
to eliminate the dependence on the hole momentum, by an average
procedure (see Ref. \cite{kn:ba95}), as follows

\begin{equation}
\Sigma^{PQ(R)P} (\bold{Q}, \nu) \equiv
\frac{1}{ \frac{4}{3} \ \pi} \ \int d^3 h \
\Sigma^{PQ(R)P} (\bold{Q}, \nu, \bold{h})
\end{equation}

\newpage

\section*{APPENDIX B}

In this appendix we show explicit expressions for
exchange contributions to the
structure function. Direct contributions
can be found in Ref. \cite{kn:ba95}.

Let us first consider the $S_{PP}$ channel.
Graph $SE2E$ of Fig. 1 (see Eq. (\ref{eq:spplise}))
is given by,

\begin{eqnarray}
(S_{SE1} (\bold{Q}, \nu))_{L,T} & = &
- \frac{A}{(2 \pi)^5}\ ( \frac{f_{\pi}^2}{4 \pi \hbar c} \ )^2
\frac{3 (m c^2)^3}{2 (\hbar c \mu_{\pi})^4} \
\int d^3 h \ \int d^3 k \ \int d^3 k' \ {\cal OV}_{(L,T),SE2E}
\nonumber \\ [5.mm]
& & \theta ( 1 - | \bold{h} | ) \theta ( | \bold{h} + \bold{Q} | -1)
\theta ( 1 - | \bold{h} + \bold{k'} | )
\theta (| \bold{h} + \bold{k} + \bold{k'} | - 1)
\nonumber \\ [5.mm]
& &
\theta ( 1 - | \bold{h} + \bold{k} | )
\theta ( | \bold{h} + \bold{k} + \bold{Q}| - 1 )
(\frac{-1}{\pi} \ Im) [
\frac{1}{ \nu - (Q^2/2 + \bold{Q.h})+i \eta} \
\nonumber \\ [5.mm]
& &
\frac{1}{ \nu - (Q^2/2 + \bold{Q.h} + \bold{k.k'} )+i \eta} \
\frac{1}{ \nu - (Q^2/2 + \bold{Q.(h+k)})+i \eta} \ ]  \ .
\label{eq:sel1}
\end{eqnarray}

Graph $SE2E'$ of Fig. 1 (see Eq. (\ref{eq:spplise})):

\begin{eqnarray}
(S_{SE2E'} (\bold{Q}, \nu))_{L,T} & = &
- \frac{A}{(2 \pi)^5}\ ( \frac{f_{\pi}^2}{4 \pi \hbar c} \ )^2
\frac{3 (m c^2)^3}{2 (\hbar c \mu_{\pi})^4} \
\int d^3 h \ \int d^3 k \ \int d^3 k' \ {\cal OV}_{(L,T),SE2E'}
\nonumber \\ [5.mm]
& & \theta ( 1 - | \bold{h} | ) \theta ( | \bold{h} + \bold{Q} | -1)
\theta ( 1 - | \bold{h} + \bold{k} | )
\theta (1 - | \bold{h} + \bold{k} + \bold{k'} + \bold{Q} | )
\nonumber \\ [5.mm]
& &
\theta ( | \bold{h} + \bold{k'} + \bold{Q} | )
\theta ( | \bold{h} + \bold{k} + \bold{Q}| - 1 )
(\frac{-1}{\pi} \ Im) [
\frac{1}{ \nu - (Q^2/2 + \bold{Q.h})+i \eta} \
\nonumber \\ [5.mm]
& &
\frac{1}{ \nu - (Q^2/2 + \bold{Q.h} - \bold{k.k'} )+i \eta} \
\frac{1}{ \nu - (Q^2/2 + \bold{Q.(h+k)})+i \eta} \ ] \ ,
\label{eq:sel2}
\end{eqnarray}
where

\begin{equation}
{\cal OV}_{L,SE2E}  = {\cal OV}_{L,SE2E'}  =
\Gamma_{\pi}^2 (k) \; \Gamma_{\pi}^2 (k') \;
\{ 3 \tilde{g}'^2 -  ( 2 (\bold{\widehat{k}.\widehat{k'}} )^2
- 1) \tilde{h}'^2  + 2 \tilde{g}' \tilde{h}' \} \ ,
\label{eq:fsel1l}
\end{equation}
and
\begin{eqnarray}
{\cal OV}_{T,SE2E} & = & {\cal OV}_{T,SE2E'} \; = \;
\nonumber \\ [5.mm]
& = &
(\frac{\hbar c k_F}{2 m c^2} \ )^2
\Gamma_{\pi}^2 (k) \; \Gamma_{\pi}^2 (k')
\{ 4 [ \bold{h.(h+k)}-\bold{(\widehat{Q}.h)(\widehat{Q}.
(h+k)}] \nonumber \\ [5.mm]
& &
[ 3 \tilde{g}'^2 - ( 2 (\bold{\widehat{k}.\widehat{k'}} )^2 - 1)
 \tilde{h}'^2  + 2 \tilde{g}' \tilde{h}'] +  \nonumber \\ [5.mm]
&  & (-3 {\mu_s}^2 + {\mu_v}^2)
[ Q^2 \tilde{g}'^2 + ((\bold{Q.k'})^2 - 2 (\bold{k.k'})
(\bold{Q.k})(\bold{Q.k'}))
\tilde{h}'^2 +  \nonumber \\ [5.mm]
&  &
(\bold{Q.\widehat{k'}})^2 \tilde{g}'(k) \tilde{h}'(k') +
(Q^2- 2 (\bold{Q.\widehat{k}})^2) \tilde{g}'(k') \tilde{h}'(k) ] -
\nonumber \\ [5.mm]
&  &
(- 3 \mu_s + \mu_v) \tilde{h}'^2 \frac{2}{Q^2} \
(\bold{\widehat{k}.\widehat{k'}} )[
(\bold{\widehat{k}.Q})\bold{\widehat{k'}}.( 2 \bold{h+k}) -
(\bold{\widehat{k'}.Q})\bold{\widehat{k}}.( 2 \bold{h+k}) ] \} \ .
\label{eq:fsqq1t}
\end{eqnarray}

First order exchange contribution to the RPA-type
correlation (see Eq. (\ref{eq:sppfrpa}) and graph
$RPA1E$ of Fig. 1):

\begin{eqnarray}
(S_{RPA1E} (\bold{Q}, \nu))_{L,T} & = &
-\frac{A}{(2 \pi)^3}\
(\frac{f_{\pi}^2}{4 \pi \hbar c } \ )
\frac{3 (m c^2)^2}{2 (\hbar c \mu_{\pi})^2 \hbar c k_F}\
\int d^3 h \ \int d^3 k \
{\cal OV}_{(L,T),RPA1E}
\nonumber \\ [5.mm]
& & \theta ( 1 - h)
\theta (| \bold{h} + \bold{Q} | - 1)
\theta (1 - | \bold{h} + \bold{k} | )
\theta (| \bold{h} + \bold{k} + \bold{Q} | - 1) \nonumber \\ [5.mm]
& &
(- \frac{1}{ \pi} \ Im )
\{ (\frac{1}
{2 \nu - (Q^2 + 2 \bold{h.Q}) + i \eta} \
 \, - \,
\frac{1}
{2 \nu + (Q^2 + 2 \bold{h.Q}) } \ )
\nonumber \\ [5.mm]
& &
(\frac{1}
{2 \nu - (Q^2 + 2 \bold{(h+k).Q} ) + i \eta} \
 \, - \,
\frac{1}
{2 \nu + (Q^2 + 2 \bold{(h+k).Q}) } \ ) \} \ ,
\label{eq:se1}
\end{eqnarray}
where
\begin{equation}
{\cal OV}_{(L),RPA1E}  =
\Gamma_{\pi}^2 (k)
( 3 \tilde{g}' + \tilde{h}' )
\end{equation}
and
\begin{eqnarray}
{\cal OV}_{(T),RPA1E} & = &
(\frac{\hbar c k_F}{2 m c^2} \ )^2
\Gamma_{\pi}^2 (k)
 \{ [ 3 \tilde{g}' + \tilde{h}'] \;
4 \; [(\bold{h.(h+k)} - (\bold{Q.h}) \bold{Q.(h+k)}]/Q^2] \; + \;
\nonumber \\ [5.mm]
& &
(-3 {\mu_s}^2 + {\mu_v}^2) Q^2
( \tilde{g}' + \tilde{h}'(\bold{\widehat{k} . \widehat{Q}})^2  \} \ .
\end{eqnarray}

Graph $RPA2E$ of Fig. 1 (see Eq. (\ref{eq:sppsrpa})):

\begin{eqnarray}
(S_{RPA2E} (\bold{Q}, \nu))_T & = &
-\frac{A}{(2 \pi)^5}\
(\frac{f_{\pi}^2}{4 \pi \hbar c } \ )^2
\frac{3 (m c^2)^3}{(\hbar c \mu_{\pi})^4}\
(- \frac{1}{ \pi} \ Im ) \{ {\cal L} (\bold{Q}, \nu)
\nonumber \\ [5.mm]
& & \int d^3 h \ \int d^3 k \  {\cal OV}_{(L,T),RPA2E}
\theta ( 1 - h)
\theta (| \bold{h} + \bold{Q} | - 1)
\theta (1 - | \bold{h} + \bold{k} | )
\nonumber \\ [5.mm]
& &
\theta (| \bold{h} + \bold{k} + \bold{Q} | - 1)
(\frac{1}
{2 \nu - (Q^2 + 2 \bold{h.Q}) + i \eta} \
 \, - \,
\frac{1}
{2 \nu + (Q^2 + 2 \bold{h.Q}) } \ )
\nonumber \\ [5.mm]
& &
(\frac{1}
{2 \nu - (Q^2 + 2 \bold{(h+k).Q} ) + i \eta} \
 \, - \,
\frac{1}
{2 \nu + (Q^2 + 2 \bold{(h+k).Q}) } \ ) \} \ ,
\label{eq:sed2}
\end{eqnarray}
with
\begin{eqnarray}
{\cal L} (\bold{Q}, \nu) & = & \int d^3 p \;
\theta( |\bold{p+Q}/2| -1) \;
\theta(1-|\bold{p-Q}/2|)
\nonumber \\ [5.mm]
& &
(\frac{1}{2 \nu - Q^2 - 2 \bold{Q.p} + i \eta} \ \, - \,
\frac{1}{2 \nu + Q^2 + 2 \bold{Q.p}} \ ) \ .
\end{eqnarray}
and
\begin{eqnarray}
{\cal OV}_{(T),RPA2E} & = &
\Gamma_{\pi}^2 (Q) \Gamma_{\pi}^2 (k) Q^2
4 \tilde{g}' {\mu_v}^2 (
\tilde{g}' + \tilde{h}'(\bold{\widehat{k} . \widehat{Q}})^2  ) \ .
\end{eqnarray}
(Note that for the present interaction the longitudinal
contribution is zero).

Graph $RPA2E'$ of Fig. 1 (see Eq. (\ref{eq:sppsrpa})):

\begin{eqnarray}
(S_{RPA2E'} (\bold{Q}, \nu))_{L,T} & = &
\frac{A}{(2 \pi)^5}\
(\frac{f_{\pi}^2}{4 \pi \hbar c } \ )^2
\frac{6 m c^2 k_F^2}{(\hbar c \mu_{\pi}^2)^2}\
\int d^3 h \ \int d^3 k \ \int d^3 k' \
{\cal OV}_{(L,T),RPA2E'}
\nonumber \\ [5.mm]
& &
\theta ( 1 - h)
\theta (| \bold{h} + \bold{Q} | - 1)
\theta (1 - | \bold{h} + \bold{k} | )
\nonumber \\ [5.mm]
& &
\theta (| \bold{h} + \bold{k} + \bold{Q} | - 1)
\theta (1 - | \bold{h} + \bold{k} + \bold{k'} | )
\theta (| \bold{h} + \bold{k} + \bold{k'} + \bold{Q} | - 1)
\nonumber \\ [5.mm]
& &
(- \frac{1}{ \pi} \ Im ) \{
(\frac{1}
{2 \nu - (Q^2 + 2 \bold{h.Q}) + i \eta} \
 \, - \,
\frac{1}
{2 \nu + (Q^2 + 2 \bold{h.Q}) } \ )
\nonumber \\ [5.mm]
& &
(\frac{1}
{2 \nu - (Q^2 + 2 \bold{(h+k).Q} ) + i \eta} \
 \, - \,
\frac{1}
{2 \nu + (Q^2 + 2 \bold{(h+k).Q} ) } \ )
\nonumber \\ [5.mm]
& &
(\frac{1}
{2 \nu - (Q^2 + 2 \bold{(h+k+k').Q} ) + i \eta} \
 \, - \,
\frac{1}
{2 \nu + (Q^2 + 2 \bold{(h+k+k').Q} ) } \ ) \} \ , 
\nonumber \\
& &
\label{eq:se2}
\end{eqnarray}
where
\begin{eqnarray}
{\cal OV}_{(L),RPA2E'} & = &
\Gamma_{\pi}^2 (k) \, \Gamma_{\pi}^2 (k')
 10 [ 9 (\tilde{g}')^2 + \tilde{h}'^2 + 6 \tilde{g}' \tilde{h}']
\end{eqnarray}
and
\begin{eqnarray}
{\cal OV}_{(T),RPA2E'} & = &
(\frac{\hbar c k_F}{2 m c^2} \ )^2
\Gamma_{\pi}^2 (k) \, \Gamma_{\pi}^2 (k')
\; \{ 40 [ 9 (\tilde{g}')^2 + \tilde{h}'^2 + 6 \tilde{g}' \tilde{h}'] \;
\nonumber \\ [5.mm]
& &
[(\bold{h.(h+k+k')} - (\bold{Q.h}) \bold{Q.(h+k+k')}]/Q^2] \; + \;
\nonumber \\ [5.mm]
& &
\frac{9 {\mu_s}^2 + {\mu_v}^2}{2} \ \{ Q^2
[ (\tilde{g}'_2)^2 + \tilde{h}'^2
( 2 (\bold{\widehat{k}.\widehat{k'}})^2 -1 ) ] \; + \;
\nonumber \\ [5.mm]
& &
\tilde{h}'^2 [ (\bold{Q.\widehat{k}})^2 \, + \,
(\bold{Q.\widehat{k'}})^2 \, - \, 2 (\bold{Q.\widehat{k}})
(\bold{Q.\widehat{k'}}) (\bold{\widehat{k}.\widehat{k'}})] \; + \;
\tilde{g}'_2 \tilde{h}' [
(\bold{Q.\widehat{k}})^2 \, + \,
(\bold{Q.\widehat{k'}})^2 ] \} \ .
\nonumber \\
& &
\end{eqnarray}

Going now to the $S_{QP}$ and $S_{QQ}$-channel
(see Eqs. (\ref{eq:sqp3}) and (\ref{eq:sqq3}));
we have for graph $SQPE$ of Fig. 2 :

\begin{eqnarray}
(S_{SQPE} (\bold{Q}, \nu))_{L,T} & = &
\frac{A}{(2 \pi)^5}\ ( \frac{f_{\pi}^2}{4 \pi \hbar c} \ )^2
\frac{3 (m c^2)^3}{4 (\hbar c \mu_{\pi})^4} \
\int d^3 h \ \int d^3 k \ \int d^3 k' \ {\cal OV}_{(L,T),SQPE}
\nonumber \\ [5.mm]
& & \theta ( 1 - | \bold{h} | ) \theta ( | \bold{h} - \bold{k} | -1)
\theta ( | \bold{h} - \bold{k} + \bold{Q} | - 1)
\theta (| \bold{h} - \bold{k'} + \bold{Q} | - 1)
\nonumber \\ [5.mm]
& &
\theta ( | \bold{h} + \bold{Q} | - 1 )
\theta ( 1 - | \bold{h} - \bold{k} - \bold{k'} + \bold{Q}| )
\frac{1}{ \bold{k.(k'-Q)}} \
\nonumber \\ [5.mm]
& &
(\frac{-1}{\pi} \ Im) [
\frac{1}{ \nu - (Q^2/2 + \bold{Q.h} - \bold{k.k'} )+i \eta} \
\frac{1}{ \nu - (Q^2/2 + \bold{Q.h)})+i \eta} \ ] \ .
\nonumber \\
& &
\label{eq:sqp1}
\end{eqnarray}

Graph $SQPE'$ of Fig. 2:

\begin{eqnarray}
(S_{SQPE'} (\bold{Q}, \nu))_{L,T} & = &
 \frac{A}{(2 \pi)^5}\ ( \frac{f_{\pi}^2}{4 \pi \hbar c} \ )^2
\frac{3 (m c^2)^3}{4 (\hbar c \mu_{\pi})^4} \
\int d^3 h \ \int d^3 k \ \int d^3 k' \ {\cal OV}_{(L,T),SQPE'}
\nonumber \\ [5.mm]
& & \theta ( 1 - | \bold{h} | ) \theta ( | \bold{h} - \bold{k} | -1)
\theta ( | \bold{h} - \bold{k'} + \bold{Q} | - 1)
\theta (| \bold{h} + \bold{Q} | - 1)
\nonumber \\ [5.mm]
& &
\theta ( | \bold{h} - \bold{k'} | - 1 )
\theta ( 1 - | \bold{h} - \bold{k} - \bold{k'}| )
\frac{1}{ \bold{k.k'}} \
\nonumber \\ [5.mm]
& &
(\frac{-1}{\pi} \ Im) [
\frac{1}{ \nu - (Q^2/2 + \bold{Q.h} - \bold{k'.(k+Q)} )+i \eta} \
\frac{1}{ \nu - (Q^2/2 + \bold{Q.h)})+i \eta} \ ] \ ,
\nonumber \\
& &
\label{eq:sqpee}
\end{eqnarray}
where

\begin{equation}
{\cal OV}_{L,SQPE}  = {\cal OV}_{L,SE2E}\ ,
\label{eq:fsqp1e}
\end{equation}

\begin{equation}
{\cal OV}_{T,SQPE}  =  {\cal OV}_{T,SE2E}\ ,
\label{eq:fsqp1t}
\end{equation}

\begin{eqnarray}
{\cal OV}_{T,SQPE'} & = &
(\frac{\hbar c k_F}{2 m c^2} \ )^2
\Gamma_{\pi}^2 (k) \; \Gamma_{\pi}^2 (k')
\{ 4 [ \bold{h.(h+k')}-\bold{(\widehat{Q}.h)(\widehat{Q}.
(h+k')}] \nonumber \\ [5.mm]
& &
[ 3 \tilde{g}'^2 - ( 2 (\bold{\widehat{k}.\widehat{k'}} )^2 - 1)
 \tilde{h}'^2  + 2 \tilde{g}' \tilde{h}'] +  \nonumber \\ [5.mm]
&  & (-3 {\mu_s}^2 + {\mu_v}^2)
[ Q^2 \tilde{g}'^2 + ((\bold{Q.k'})^2 - 2 (\bold{k.k'})
(\bold{Q.k})(\bold{Q.k'}))
\tilde{h}'^2 +  \nonumber \\ [5.mm]
&  &
(\bold{Q.\widehat{k'}})^2 \tilde{g}'(k) \tilde{h}'(k') +
(Q^2- 2 (\bold{Q.\widehat{k}})^2) \tilde{g}'(k') \tilde{h}'(k) ] -
\nonumber \\ [5.mm]
&  &
(- 3 \mu_s + \mu_v) \tilde{h}'^2 \frac{2}{Q^2} \
(\bold{\widehat{k}.\widehat{k'}} )[
(\bold{\widehat{k}.Q})\bold{\widehat{k'}}.( 2 \bold{h+k'}) -
(\bold{\widehat{k'}.Q})\bold{\widehat{k}}.( 2 \bold{k+k'}) ] \}\ .
\nonumber \\
& &
\label{eq:fsqp2t}
\end{eqnarray}

Graph $SQPD'$ of Fig. 2:

\begin{eqnarray}
(S_{QPD'} (\bold{Q}, \nu))_{L,T} & = &
- \frac{A}{(2 \pi)^5}\ ( \frac{f_{\pi}^2}{4 \pi \hbar c} \ )^2
\frac{3 (m c^2)^3}{2 (\hbar c \mu_{\pi})^4} \
\int d^3 h \ \int d^3 h' \ \int d^3 k \ {\cal OV}_{(L,T),SQPD'}
\nonumber \\ [5.mm]
& & \theta ( 1 - | \bold{h} | ) \theta ( | \bold{h} + \bold{Q} | -1)
\theta ( | \bold{h} - \bold{k} | -1 )
\theta (| \bold{h'} + \bold{k} + \bold{Q} | - 1)
\nonumber \\ [5.mm]
& &
\theta (| \bold{h'} + \bold{k} | - 1)
\theta ( 1 - | \bold{h'} | ) \frac{1}{k^2-\bold{k.h}+\bold{k.h'}} \
(\frac{-1}{\pi} \ Im)
\nonumber \\ [5.mm]
& &
\frac{1}{ \nu - (Q^2/2 + \bold{Q.h})+i \eta} \
\frac{1}{ \nu - (k^2+Q^2/2 + \bold{k.(h'-h)}+\bold{Q.(k+h')})+i \eta}\ . 
\nonumber \\
& &
\label{eq:sqpdd}
\end{eqnarray}

Graph $SQQ3D'$ of Fig. 3:

\begin{eqnarray}
(S_{QQ3D'} (\bold{Q}, \nu))_{L,T} & = &
\frac{A}{(2 \pi)^5}\ ( \frac{f_{\pi}^2}{4 \pi \hbar c} \ )^2
\frac{3 (m c^2)^3}{2 (\hbar c \mu_{\pi})^4} \
\int d^3 h \ \int d^3 h' \ \int d^3 k \ {\cal OV}_{(L,T),SQQ3D'}
\nonumber \\ [5.mm]
& & \theta ( 1 - | \bold{h} | ) \theta ( | \bold{h} - \bold{k} | -1)
\theta ( 1 - | \bold{h'} + \bold{Q} | )
\theta (| \bold{h} - \bold{k} + \bold{Q} | - 1)
\nonumber \\ [5.mm]
& &
\theta (| \bold{h'} + \bold{k} | - 1)
\theta ( 1 - | \bold{h'} | )
\nonumber \\ [5.mm]
& &
\frac{1}{k^2-\bold{k.h}+\bold{k.h'}} \
\frac{1}{k^2+\bold{k.(h'-h)} +\bold{Q.(h-h'-k)}} \
\nonumber \\ [5.mm]
& &
(\frac{-1}{\pi} \ Im)
\frac{1}{ \nu - (k^2+Q^2/2 + \bold{k.(h'-h)}+\bold{Q.(h-k)})+i \eta} \ .
\label{eq:sqq3dd}
\end{eqnarray}
In Eqs. (\ref{eq:sqpdd})-(\ref{eq:sqq3dd}),
we have used, $$\bold{k'}=\bold{k-Q}$$
and

\begin{equation}
{\cal OV}_{L,SQPD'}  = {\cal OV}_{L,SQQ3D'}  =
\Gamma_{\pi}^2 (k) \; \Gamma_{\pi}^2 (k') \;
5 \{ 3 \tilde{g}'^2 +  (\bold{\widehat{k}.\widehat{k'}} )^2
 \tilde{h}'^2  + 2 \tilde{g}' \tilde{h}' \} \ ,
\label{eq:fsqp3l}
\end{equation}

\begin{eqnarray}
{\cal OV}_{T,SQPD'} & = &
(\frac{\hbar c k_F}{2 m c^2} \ )^2
\Gamma_{\pi}^2 (k) \; \Gamma_{\pi}^2 (k')
\nonumber \\ [5.mm]
& &
\{ 20 [ \bold{h.(h'+k)}-\bold{(\widehat{Q}.h)(\widehat{Q}.
(h'+k)}][
3 \tilde{g}'^2 +  (\bold{\widehat{k}.\widehat{k'}} )^2
 \tilde{h}'^2  + 2 \tilde{g}' \tilde{h}'] +  \nonumber \\ [5.mm]
&  & (3 {\mu_s}^2 + 2 {\mu_v}^2)
[4 Q^2 \tilde{g}'^2 + (\bold{Q} \times (\bold{k} \times \bold{k'}))^2
\tilde{h}'^2 +  \nonumber \\ [5.mm]
&  &
(Q^2+(\bold{Q.\widehat{k'}})^2) \tilde{g}'(k) \tilde{h}'(k') +
(Q^2+(\bold{Q.\widehat{k}})^2) \tilde{g}'(k') \tilde{h}'(k) ] -
\nonumber \\ [5.mm]
&  &
(3 \mu_s + 2 \mu_v) \tilde{h}'^2 \frac{2}{Q^2} \
(\bold{\widehat{k}.\widehat{k'}} )[
(\bold{\widehat{k}.Q})(\bold{\widehat{k'}.(h+h'+k)}) -
(\bold{\widehat{k'}.Q})(\bold{\widehat{k}.(h+h'+k')}) ] \} \ ,
\nonumber \\
& &
\label{eq:fsqp3t}
\end{eqnarray}
and
\begin{eqnarray}
{\cal OV}_{T,SQQ3D'} & = &
(\frac{\hbar c k_F}{2 m c^2} \ )^2
\Gamma_{\pi}^2 (k) \; \Gamma_{\pi}^2 (k')
\nonumber \\ [5.mm]
& &
\{ 20 [ \bold{h'.(h-k)}-\bold{(\widehat{Q}.h')(\widehat{Q}.
(h-k)}][
3 \tilde{g}'^2 +  (\bold{\widehat{k}.\widehat{k'}} )^2
 \tilde{h}'^2  + 2 \tilde{g}' \tilde{h}'] +  \nonumber \\ [5.mm]
&  & (3 {\mu_s}^2 + 2 {\mu_v}^2)
[4 Q^2 \tilde{g}'^2 + (\bold{Q} \times (\bold{k} \times \bold{k'}))^2
\tilde{h}'^2 +  \nonumber \\ [5.mm]
&  &
(Q^2+(\bold{Q.\widehat{k'}})^2) \tilde{g}'(k) \tilde{h}'(k') +
(Q^2+(\bold{Q.\widehat{k}})^2) \tilde{g}'(k') \tilde{h}'(k) ] -
\nonumber \\ [5.mm]
&  &
(3 \mu_s + 2 \mu_v) \tilde{h}'^2 \frac{2}{Q^2} \
(\bold{\widehat{k}.\widehat{k'}} )[
(\bold{\widehat{k}.Q})(\bold{\widehat{k'}.(h+h'-k)}) -
(\bold{\widehat{k'}.Q})(\bold{\widehat{k}.(h+h'-k')}) ] \} \ .
\nonumber \\
& &
\label{eq:fsqq3t}
\end{eqnarray}

Graph $SQQ1E$ of Fig. 3:

\begin{eqnarray}
(S_{SQQ1E} (\bold{Q}, \nu))_{L,T} & = &
- \frac{A}{(2 \pi)^5}\ ( \frac{f_{\pi}^2}{4 \pi \hbar c} \ )^2
\frac{9 (m c^2)^3}{4 (\hbar c \mu_{\pi})^4} \
\int d^3 p \ \int d^3 k \ \int d^3 k' \
\nonumber \\ [5.mm]
& & {\cal OV}_{(L,T),SQQ1E}
\theta ( | \bold{p} | - 1 ) \theta ( 1 - | \bold{p} + \bold{k} | )
\theta ( | \bold{p} + \bold{Q} | - 1)
\nonumber \\ [5.mm]
& & \theta ( 1 - | \bold{h} + \bold{k'} | )
\theta ( | \bold{p} + \bold{k} + \bold{k'}| - 1)
\nonumber \\ [5.mm]
& &
\frac{1}{ (\bold{k.k'})^2 } \
(\frac{-1}{\pi} \ Im) [
\frac{1}{ \nu - (Q^2/2 + \bold{Q.p} + \bold{k'.k} )+i \eta} \ ] \ .
\label{eq:sqqp}
\end{eqnarray}

Finally, graph $SQQ2E$ of Fig. 3:

\begin{eqnarray}
(S_{SQQ2E} (\bold{Q}, \nu))_{L,T} & = &
- \frac{A}{(2 \pi)^5}\ ( \frac{f_{\pi}^2}{4 \pi \hbar c} \ )^2
\frac{9 (m c^2)^3}{4 (\hbar c \mu_{\pi})^4} \
\int d^3 h \ \int d^3 k \ \int d^3 k' \
\nonumber \\ [5.mm]
& & {\cal OV}_{(L,T),SQQ2E}
\theta ( 1 - | \bold{h} | ) \theta ( | \bold{h} - \bold{k} | -1)
\theta ( | \bold{h} - \bold{k'} | - 1)
\nonumber \\ [5.mm]
& & \theta ( 1 - | \bold{h} - \bold{Q} | )
\theta ( 1 - | \bold{h} - \bold{k} - \bold{k'}| )
\nonumber \\ [5.mm]
& &
\frac{1}{ (\bold{k.k'})^2 } \
(\frac{-1}{\pi} \ Im) [
\frac{1}{ \nu - (Q^2/2 + \bold{Q.h} - \bold{k'.k} )+i \eta} \ ] \ ,
\label{eq:sqqh}
\end{eqnarray}
where

\begin{equation}
{\cal OV}_{L,SQQ1E}  = {\cal OV}_{L,SQQ2E}
= {\cal OV}_{L,SE2E} \ ,
\label{eq:fsqqle}
\end{equation}

\begin{eqnarray}
{\cal OV}_{T,SQQ1E}  & = &
(\frac{\hbar c k_F}{2 m c^2} \ )^2
\Gamma_{\pi}^2 (k) \; \Gamma_{\pi}^2 (k')
\{ [ \bold{p}^2-(\bold{\widehat{Q}.p})^2 ]
+ Q^2 ({\mu_s}^2 + {\mu_v}^2))/4 \}
\nonumber \\ [5.mm]
& &
[ 3 \tilde{g}'^2 - ( 2 (\bold{\widehat{k}.\widehat{k'}} )^2 - 1)
 \tilde{h}'^2  + 2 \tilde{g}' \tilde{h}'] \ ,
\label{eq:fsqqpt}
\end{eqnarray}
and

\begin{eqnarray}
{\cal OV}_{T,SQQ2E}  & = &
(\frac{\hbar c k_F}{2 m c^2} \ )^2
\Gamma_{\pi}^2 (k) \; \Gamma_{\pi}^2 (k')
\{ [ \bold{h}^2-(\bold{\widehat{Q}.h})^2 ]
+ Q^2 ({\mu_s}^2 + {\mu_v}^2))/4 \}
\nonumber \\ [5.mm]
& &
[ 3 \tilde{g}'^2 - ( 2 (\bold{\widehat{k}.\widehat{k'}} )^2 - 1)
 \tilde{h}'^2  + 2 \tilde{g}' \tilde{h}']  \ .
\label{eq:fsqqht}
\end{eqnarray}

\begin{table}
\caption{Free and Self-energy contributions to the
Longitudinal and Transverse structure function (multiplied by $10^5$). 
All results
are for nuclear matter at momentum transfer $q=410 {\rm MeV/c}$
in units of MeV$^{-1}$ fm$^{-3}$.
The first column represents the energy transfer in MeV. Column $Lind.$
represents the free structure function. Columns $SE13$ give
the direct (D) and exchange (E) contribution to the diagonal
part of the self-energy up to infinite order. Their first
contributions
are the graphs $SE1_D$, $SE3_D$, $SE1_E$,  and $SE3_E$ of
Fig. 1.
Columns $SE2_D$, $SE2_E$ and $SE2_{E'}$
are the non-diagonal self-energy contributions to the structure
function as shown in Fig. 1. The last column is the sum of all
these contributions given by Eq. (\protect\ref{eq:spplise}).}

\begin{tabular}{cccccccc}	     
$long.$ & & & & & & \\ \hline
$\hbar \omega$& Lind.& $SE13_D$  & $SE13_E$ & $SE2_D$ & $SE2_E$ & $SE2_{E'}$ & $S^{Lind.+SE.}$ \\ \hline
 50. &	 38.729  &   -1.269  &	 0.715	&-1.570   &  0.185  &	0.071	 &  36.862     \\
100. &	 46.106  &   -4.390  &	 0.599	&-0.559   &  0.374  &	0.159	 &  42.289     \\
150. &	 41.357  &   -5.949  &	 0.397	& 1.085   & -0.096  &  -0.050	 &  36.743     \\
200. &	 24.481  &   -3.250  &	 0.111	& 2.219   & -0.045  &  -0.027	 &  23.490     \\
250. &	  0.000  &    2.971  &	-0.127	& 0.521   & -0.004  &  -0.005	 &   3.357     \\ \hline\hline
$trans.$ & & & & & & \\ \hline
 50. &	 60.380 &  -1.938   & ~1.112  & -3.205	& ~0.037  & ~0.181 &  56.566	  \\
100. &	 72.299 &  -6.943   & ~0.946  & -0.604	& ~0.119  & ~0.004 &  65.821	  \\
150. &	 64.610 &  -9.354   & ~0.621  & ~2.575	& -0.071  & -0.232 &  58.149	  \\
200. &	 37.739 &  -4.948   & ~0.164  & ~5.337	& -0.082  & -0.357 &  37.853	  \\
250. &	 ~0.000 &  ~4.587   & -0.195  & ~0.726	& -0.006  & -0.079 &  ~5.032	  \\ 
\end{tabular}
 \end{table}

\begin{table}
\caption{
RPA-type contributions to the Transverse structure function 
(multiplied by $10^5$) in units of MeV$^{-1}$ fm$^{-3}$ 
for nuclear matter at momentum transfer $q=410$ MeV/c.
Columns $RPA1_D$ ($RPA1_E$) and $RPA2_D$ ($RPA2_{(E+E')}$)
are the first and second order direct (exchange)-part
to the RPA response, respectively. The notation is
the same as in Fig. 1. Note that in that figure only
forward-going contributions are shown while the present
results contain both forward and backward-going contributions.
Column $RPA12_{D+E}$ is the sum of all first and second order
contributions (given by Eqs. (\protect\ref{eq:sppfrpa}) and
(\protect\ref{eq:sppsrpa})).
Finally, column $RPA_{ant.}$ is the result for a full antisymmetric
RPA using the formalism given in
Ref. \protect\cite{kn:ba96}.}

\begin{tabular}{ccccccc}	     
$\hbar \omega$ & & & & & $RPA12_{D+E}$ & $RPA_{ant.}$	\\
\cline{2-6}
 & $RPA1_D$ & $RPA1_E$ & $RPA2_D$ & $RPA2_{(E+E')}$ &  &  \\ \hline
 50. &-30.233 &~8.844 &~9.028 & -2.912 &-15.273 &-19.527\\
100. &-13.801 &~4.209 &-3.098 & ~0.914 &-11.775 &~-9.988\\
150. &~~5.122 &-1.132 &-3.441 & ~1.204 &~~1.753 &~~3.307\\
200. &~10.354 &-3.206 &~1.468 & -0.466 &~~8.150 &~~7.186\\
250. &~~0.000 &~0.000 &~0.000 & ~0.000 &~~0.000 &~~0.000\\
\end{tabular}
\end{table}

\begin{table}
\caption{
Longitudinal and Transverse $S_{QP}$-type structure function
(multiplied by $10^5$) in units of MeV$^{-1}$ fm$^{-3}$ 
for nuclear matter at momentum transfer $q=410$ MeV/c.
The notation $SQP_D$ to $SQP_{D'}$ is the same as in Fig. 2.
Column $SQP_{ant}$ is the sum of all contributions.
}

\begin{tabular}{cccccc} 	   
$long.$ & & & & & \\ \hline
$\hbar \omega$	& $SQP_D$ & $SQP_E$ & $SQP_{E'}$ & $SQP_{D'}$ & $SQP_{ant}$ \\ \hline
 50. &	 4.950 & -0.003   &  -0.077   &   1.101  &   5.971   \\
100. &	 4.136 & -0.016   &  -0.110   &   1.142  &   5.152   \\
150. &	 3.116 & -0.055   &  -0.075   &   0.589  &   3.575   \\
200. &	 1.598 & -0.037   &  -0.007   &   0.234  &   1.789   \\
250. &	-1.284 &  0.004   &   0.022   &  -0.058  &  -1.316   \\ \hline\hline
$trans.$ & & & & & \\ \hline
 50.  &   7.469  &  -0.604  &  -0.101  &  1.324   &  8.089   \\
100.  &   5.412  &  -0.672  &  -0.144  &  1.374   &  5.969   \\
150.  &   3.656  &  -0.436  &  -0.098  &  0.928   &  4.049   \\
200.  &   0.906  &  -0.001  &  -0.009  &  0.491   &  1.387   \\
250.  &  -3.652  &   0.341  &	0.029  & -0.073   & -3.356   \\ 
\end{tabular}
\end{table}

\begin{table}
\caption{
Longitudinal and Transverse $S_{QQ}$-type structure function.
The notation $SQQ1_D$ to $SQQ3_{D}$ is the same as in Fig. 3.
$SQQ3_{Etot.}$ is the sum of $SQQ3_{E}$, $SQQ3_{E'}$
and $SQQ3_{D'}$ from the same figure.
Column $SQQ_{ant}$ is the sum of all contributions.
}

\begin{tabular}{cccccccc}	      
$long.$ & & & & & \\ \hline
$\hbar \omega$	& $SQQ1_D$ & $SQQ1_E$ & $SQQ2_D$ & $SQQ2_E$ & $SQQ3_D$ & $SQQ3_{Etot.}$ & $SQQ_{ant}$ \\ \hline
 50. &	 4.692 &  -0.042 &   2.455 &  -0.074 & -0.552  &  0.041  &   6.520   \\
100. &	 8.535 &  -0.327 &   1.428 &  -0.046 & -0.629  &  0.042  &   9.003   \\
150. &	 9.807 &  -0.344 &   0.661 &  -0.019 & -0.299  &  0.029  &   9.834   \\
200. &	10.426 &  -0.278 &   0.310 &  -0.008 & -0.220  &  0.020  &  10.249   \\
250. &	10.342 &  -0.167 &   0.047 &  -0.003 & -0.113  &  0.011  &  10.118   \\ \hline\hline
$trans.$ & & & & & \\ \hline
 50. &	 6.064 &  -0.051 &   4.413 &  -0.062 & -1.614  &  0.012  &  8.762    \\
100. &	12.673 &  -0.307 &   2.493 &  -0.027 & -2.230  &  0.017  & 12.617    \\
150. &	15.153 &  -0.313 &   1.179 &  -0.007 & -1.848  &  0.014  & 14.178    \\
200. &	17.352 &  -0.263 &   0.559 &  -0.003 & -0.467  &  0.012  & 17.188    \\
250. &	16.551 &  -0.167 &   0.090 &  -0.001 & -0.271  &  0.006  & 16.208    \\ 
\end{tabular}
\end{table}

\begin{table}
\caption{
Comparison between our results and the corresponding direct
values calculated with
a modified value of the $g'$-Landau Migdal parameter ($g'=0.5$) to
partially reproduce the exchange contributions. Our
results are given by $SPP_{ant}$, $SQP_{ant}$ and $SQQ_{ant}$ where
a value $g'=0.7$ was employed with the explicit inclusion of exchange
graphs.}

\begin{tabular}{ccccccc}       
$long.$ 	 &    &  &  &  &  &		  \\ \hline
$\hbar \omega$	 & $SPP_{ant}$ & $SPP_{dir, g'=.5}$ & $SQP_{ant}$ & $SQP_{dir, g'=.5}$& $SQQ_{ant}$ & $SQQ_{dir, g'=.5}$    \\ \hline
  50.		 & 36.862   &  36.409	 &  ~5.971   &	~5.034	   &  ~6.520   &    2.529   \\
 100.		 & 42.289   &  43.256	 &  ~5.152   &	~5.442	   &  ~9.003   &    4.995   \\
 150.		 & 36.743   &  36.430	 &  ~3.575   &	~4.312	   &  ~9.834   &    5.971   \\
 200.		 & 23.490   &  19.059	 &  ~1.789   &	~0.966	   &  10.249   &    6.291   \\
 250.		 & ~3.357   &  ~2.393	 &  -1.316   &	-1.336	   &  10.118   &    6.015   \\	 \hline\hline
$trans.$	 &    &   &  &	&  &		  \\ \hline
$\hbar \omega$	 & $SPP_{ant}$ & $SPP_{dir, g'=.5}$& $SQP_{ant}$ & $SQP_{dir, g'=.5}$& $SQQ_{ant}$ & $SQQ_{dir, g'=.5}$    \\ \hline
  50.		 &  37.039    &  47.428    & ~7.109   &  ~5.533    & ~8.762    &  ~5.049    \\
 100.		 &  55.833    &  63.337    & ~5.362   &  ~5.719    & 12.617    &  ~9.092    \\
 150.		 &  61.456    &  59.066    & ~4.049   &  ~3.864    & 14.178    &  10.814    \\
 200.		 &  45.039    &  33.882    & ~1.727   &  ~1.392    & 17.188    &  11.586    \\
 250.		 &  ~5.032    &  ~3.719    & -3.356   &  -1.883    & 16.208    &  10.800    \\	
\end{tabular}
\end{table}

\begin{figure}
\caption{Goldstone diagrams stemming from Eqs.
(\protect\ref{eq:spplise})-(\protect\ref{eq:sppsrpa}).
In every diagram the wavy lines represent the external
probe with energy momentum $({\bf q},\omega)$. The
dashed line is the residual interaction.
For simplicity we show only forward-going contributions,
where the incoming external probe
creates a particle-hole pair. In the backward-going
diagrams (not represented here) the probe  can also destroy 
a particle-hole pair.}
\end{figure}
\begin{figure}
\caption{The same as Fig. 1 but for the $S_{QP}$-channel given by Eq.
(\protect\ref{eq:sqp3}).
The action of the external probe represents
the interference between scattering (a particle or a hole)
and creating (or destroying) a particle-hole pair.}
\end{figure}
\begin{figure}
\caption{
The same as Fig. 1 but for the $S_{QQ}$-channel given by Eq.
(\protect\ref{eq:sqq3}).
The action of the external probe is to
create (or destroy) a particle-hole pair.
}
\end{figure}
\begin{figure}
   \setlength{\unitlength}{1mm}
   \begin{picture}(100,180)
   \put(25,0){\epsfxsize=12cm \epsfbox{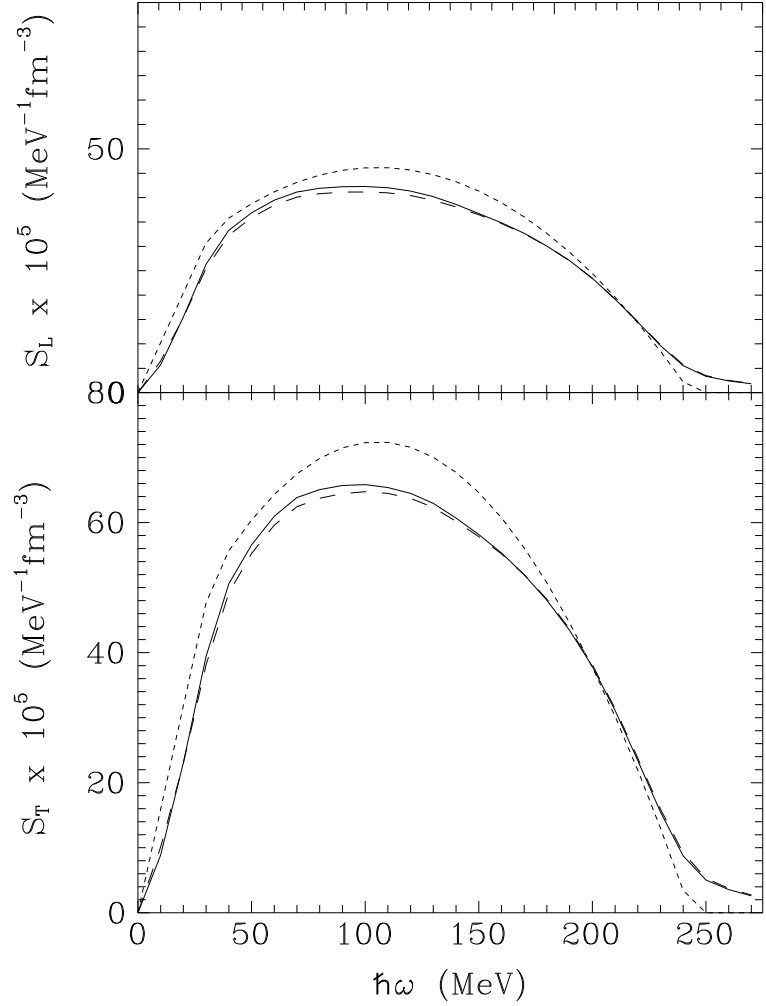}}
   \end{picture}
\caption{
Self-energy contributions to the longitudinal (upper part) and 
transverse (lower part) structure function of nuclear matter at
momentum transfer $q=410$ MeV/c. Short-dashed line: Lindhard function
using an effective mass $m^*/m{}=0.85$. Long-dashed line: effect of
adding the direct self-energy terms.
 Full line: effect of adding the direct and exchange self-energy terms.
}
\end{figure}
\begin{figure}
   \setlength{\unitlength}{1mm}
   \begin{picture}(100,180)
   \put(25,0){\epsfxsize=12cm \epsfbox{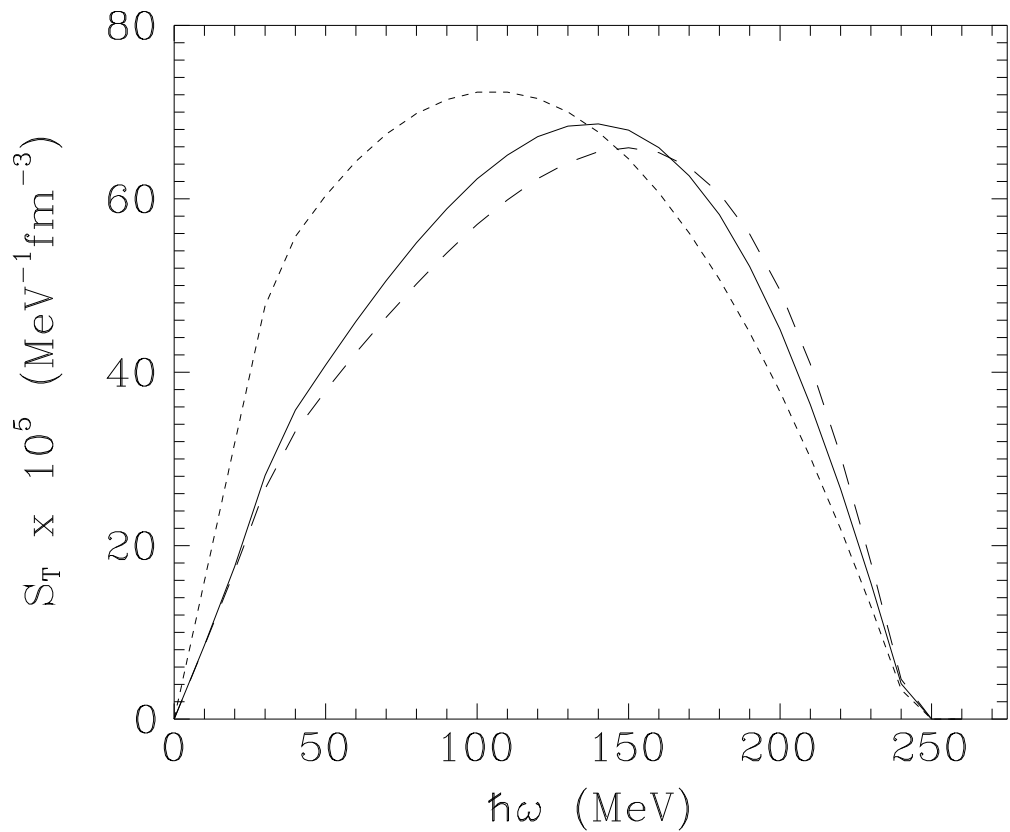}}
   \end{picture}
\caption
{RPA contributions to the transverse structure function of nuclear
matter
at $q=410$ MeV/c. Short-dashed line: Lindhard function ($m^*/m{}=0.85$).
Long-dashed line: direct ring diagrams.  Full line: full
RPA structure function including the exchange terms to all orders.
}
\end{figure}
\begin{figure}
   \setlength{\unitlength}{1mm}
   \begin{picture}(100,180)
   \put(25,0){\epsfxsize=12cm \epsfbox{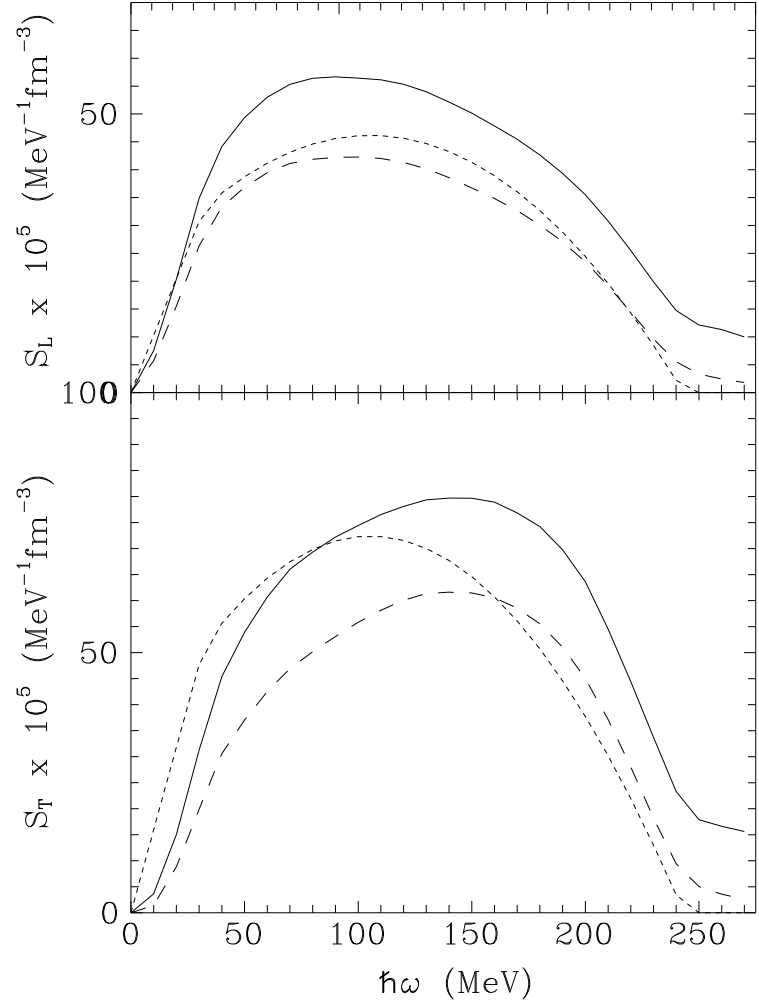}}
   \end{picture}
\caption{
Contribution of ground state correlations to the longitudinal (upper part)
and transverse (lower part) structure function of nuclear matter at momentum
transfer $q=410$ MeV/c. Short-dashed line: Lindhard function ($m^*/m{}=0.85$).
Long-dashed line: $S_{PP}$ structure function. Full line: inclusion of the 
ground state correlation diagrams to the $S_{PP}$ structure function.
}
\end{figure}

\end{document}